\documentstyle[12pt,epsfig]{article}
\begin{document}
\sloppy
\thispagestyle{empty}

\begin{flushright}
hep-ph/9902356
\end{flushright}
\vspace*{\fill}
\begin{center}
{\LARGE\bf  Spin  Effects  in Diffractive $J/\Psi$
Leptoproduction and Structure of Pomeron Coupling}\\
\vspace{2em}
\large
S.V.Goloskokov
\footnote{Email: goloskkv@thsun1.jinr.ru}
\\
\vspace{2em}
Bogoliubov Laboratory of Theoretical  Physics,\\
  Joint Institute for Nuclear Research.
 \\
 Dubna 141980, Moscow region, Russia.
\end{center}
\vspace*{\fill} %
\begin{abstract}
\noindent We calculate the cross section and the  $A_{ll}$
asymmetry of the diffractive vector meson leptoproduction for a
simple model of the pomeron coupling with the proton. It is
found that the sensitivity of the spin--dependent cross section
of the diffractive $J/\Psi$ production to the pomeron coupling
structure is rather weak. The conclusion is made that it will be
difficult to study the structure of the pomeron coupling with
the proton in future polarized diffractive experiments on the
$J/\Psi$ production.

\end{abstract}

PACS. 12.38.Bx, 13.60.Le, 13.88.+e
\vspace*{\fill}
\newpage

\section{Introduction} \label{sect1}

Study of diffractive processes at HERA has renewed interest in
investigating the nature of the pomeron. New results on the
pomeron intercept in diffractive events and information about
the pomeron partonic structure  have been obtained in H1 and
ZEUS experiments \cite{h1_zeus,h1_zeusp}. Among different
diffractive processes, which have been studied experimentally at
DESY, the vector meson production \cite{jpsi1,jpsi2} takes a
keystone place. These reactions can give information on the
gluon distribution in the nucleon at small $x$ and on the
structure of the pomeron. The diffractive $J/\Psi$ production
has a significant role in these investigations. In contrast with
the $\rho$ meson production, the $q \bar q$ exchange in the $t$
-channel is not important here and the predominant contribution
is determined by a color singlet $t$ -channel exchange
(pomeron).

The phenomenological pomeron describes the cross section
of elastic reactions at high energies and the low--$x$ behaviour
of the structure functions. From the fit of soft elastic processes
the linear soft pomeron trajectory \cite{pom_tr} was suggested
\begin{equation}
\label{pomer}
\alpha_p(t)=1+\epsilon^s+\alpha' t,
\end{equation}
with $\epsilon^s=0.08$ and $\alpha'=0.25\mbox{GeV}^{-2}$.
However, the value $\alpha_p(0)=1.12-1.2$ of the pomeron
intercept, which has been obtained at HERA \cite{h1_zeusp}, is
inconsistent with the soft pomeron with $\epsilon^s=0.08$. The
explanation of this discrepancy may be quite simple. In the soft
reactions the interaction time is large and the pomeron
rescattering effects must be important.  It has been found in
\cite{gol-p} that the rescattering effects decrease the value
of the pomeron intercept from $\epsilon \sim 0.15$ for the
hard "bare" pomeron to $\epsilon \sim 0.08$ for the soft
pomeron. Thus, in hard diffractive processes like the $J/\Psi$
production the value of $\epsilon$ should be about 0.15.

Different models of the pomeron \cite{pomeron} have been used to
study the diffractive  vector meson production
\cite{mod,cudell,rys,j-psi}. The models \cite{mod,cudell} based
on the dominant role of the pomeron contribution in these
processes reproduce the main features of the vector meson
production:  the mass, $Q^2, t$ and energy dependencies of the
cross section. In the QCD-inspired models the pomeron is modeled
by two gluons \cite{low}. It is usually assumed  that the
pomeron  couples to a proton like a $C=+1$ isoscalar photon
\cite{lansh-m} and has a simple $\gamma_{\mu}$ vertex. A more
general form of the pomeron coupling as isoscalar nucleon
current with the isoscalar Dirac and Pauli form factors have
been used in \cite{nach,klen} to study the diffractive
processes. In these approaches, the cross sections are dependent
on the pomeron coupling with the proton. Otherwise, it has been
found in \cite{rys,j-psi} that the cross-section of the
vector-meson photoproduction in the forward limit $|t|=0$ and
high $Q^2$ is proportional  to $[x G(x,\bar Q^2)]^2$. The
typical scale here is determined by $\bar Q^2=(Q^2+M_V^2)/4$
\cite{rys,j-psi} where $Q^2$ is the photon virtuality and $M_V$ is
the mass of the vector meson. For the diffractive $J/\Psi$
production the scale is large enough even for small $Q^2$ and
 perturbative calculations can be used. We see that the cross
sections of diffractive reactions are expressed, on one hand, in
terms of the pomeron coupling  and, on the other hand, through
 gluon distributions. We see, that these quantities should be
related.

The sensitivity of diffractive lepto and photoproduction to the
gluon density in the proton can give an excellent tool to test
G(x) \cite{rys}. The relation of the spin-average diffractive
production with the gluon structure function of the proton gives
way to an assumption that the longitudinal double-spin asymmetry
in such processes might be proportional to $[\Delta G/G]$
\cite{rys_dg}. Contrary to those results, in Ref. \cite{mank} it
has been found that the $A_{ll}$ asymmetry in the diffractive
vector meson  production should be zero for $|t|=0$. As a
result, this process can not be used to study $\Delta G$ of the
proton.

The value of the asymmetry  in the polarized vector meson
production in diffractive processes for nonzero $|t|$  is not
well known  now. It is very important to perform  model
calculations of the spin-dependent $J/\Psi$ leptoproduction to
obtain  quantitative estimations of spin asymmetries and their
connection with the pomeron coupling structure (see
\cite{gol_jpsi}). In this paper, we calculate the cross section
and the $A_{ll}$ asymmetry of the diffractive $J/\Psi$
leptoproduction. The cross section of the $J/\Psi$
leptoproduction can be decomposed into the leptonic and hadronic
tensors, the amplitude of the $\gamma^\star I\hspace{-1.6mm}P
\to J/\Psi$ transition amplitude and the pomeron exchange. After
describing some kinematics of the process in Sect. 2,  we
consider  the structure of the leptonic and hadronic tensors in
Sect. 3. We use a simple form of the proton coupling with the
two-gluon system which is similar to those introduced in
\cite{nach}.  In Sect. 4, we calculate the spin dependent cross
section of the $J/\Psi$ leptoproduction for the longitudinal
polarization of the initial lepton and proton.  The numerical
results for the diffractive $J/\Psi$ production at HERA and
HERMES energies is presented in Sect. 5. We finish with the
concluding remarks in Sect. 6.

\section{Kinematics of Diffractive $J/\Psi$ Leptoproduction}
\label{sect2}
Let us study the diffractive $J/\Psi$ production
\begin{equation}
\label{react}
l+p \to l+p +J/\Psi
\end{equation}
at high energies $s=(p+l)^2$ and fixed momentum transfer
$t=r_P^2=(p-p')^2$. Here $p$ and $l$ are the initial momenta of
the lepton and proton, $p'$ is the final proton momentum and
$r_P$ is a momentum carried by the pomeron. The graphs, which
describe reaction \ref{react}, are shown in Fig.\ 1 a, b.
\begin{figure}
\epsfxsize=8cm
\centerline{\epsfbox{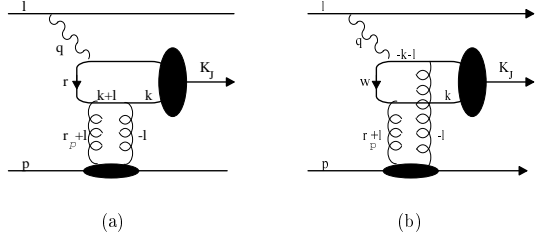}}
\caption{Feynman graphs for the diffractive vector meson
production.}
\label{fig:1}       
\end{figure}
 The
gluons from the pomeron are coupled with single and different
quarks in the $c \bar c$ loop. This ensures the gauge invariance
of the final result \cite{diehl}. The blob in the proton line
represents the proton--two--gluon coupling which comprises the
high energy $t$ channel gluon ladder of the pomeron except the
simple two-gluon exchange. The   reaction (\ref{react}) is
described in addition to $s$ and $t$ by the variables
\begin{equation}
\label{momen}
\nonumber \\
Q^2=-q^2,\quad
 y=\frac{p \cdot q}{p \cdot l},\quad x_P=\frac{q\cdot r_P}{q \cdot
p},
\end{equation}
where $Q^2$ is the photon virtuality.

The light--cone variables convenient for our calculations are
determined by $a_\pm=a_0 \pm a_z$. In these variables the scalar
production of two 4-vectors look like
$$ a \cdot
b=\frac{1}{2}(a_{+} b_{-} +a_{-} b_{+}) - \vec a_\bot \vec
b_\bot, $$
where $\vec a_\bot$ and $\vec b_\bot$ represent the
transverse parts of the momenta. We use the center of mass system
where the momenta of the initial  lepton and proton are going
along the $z$ axis and have the form
\begin{equation}
l=(p_{+},\frac{\mu^2}{p_{+}},\vec 0),\quad p=(\frac{m^2}{p_{+}},p_{+},\vec
0).
\end{equation}
Here $\mu$ and $m$ are the lepton and proton mass. The energy of
the lepton--proton system then reads as $  s \sim p_{+}^2$.
We can determine the spin vectors with positive helicity of the
lepton and the proton by
\begin{eqnarray}
\label{sp}
s_l&=&\frac{1}{\mu}(p_{+},-\frac{\mu^2}{p_{+}},\vec 0),\quad\;
s_l^2=-1,\quad  s_l \cdot l=0;\nonumber \\
s_p&=&\frac{1}{m}(-\frac{m^2}{p_{+}},p_{+},\vec 0),\quad
s_p^2=-1,\quad  s_p \cdot p=0.
\end{eqnarray}

The momenta are carried by the photon and the pomeron and can be
written as follows:
\begin{eqnarray}
q&=&(y p_{+},-\frac{Q^2}{p_{+}},\vec q_\bot),\quad
|q_\bot|=\sqrt{Q^2 (1-y)},\quad q^2=-Q^2;
\nonumber \\
r_P&=&(-\frac{|t|}{p_{+}},x_P p_{+},\vec r_\bot),\;\;
|r_\bot|=\sqrt{|t| (1-x_P)},\;\;\ r_P^2=t.
\end{eqnarray}
Thus, $y$ and $x_P$  are the fractions of the longitudinal
momenta of the lepton and proton carried by the photon and
pomeron, respectively. From the mass-shell equation for
vector--meson momentum $K_J^2=(q+r_P)^2=M_J^2$ we find that for
these reactions
\begin{equation}
\label{x_P}
x_P \sim \frac{m_J^2+Q^2+|t|}{s y}
\end{equation}
and it is small at high energies.

\section{Structure of Leptonic and Hadronic Tensors}\label{sect3}
\subsection{Leptonic Tensor}
 The
structure of the leptonic tensor is quite simple \cite{efrem}
because the lepton is a point-like object
\begin{eqnarray}
\label{lept}
{\cal L}^{\mu \nu}(s_l)&=& \sum_{spin \ s_f} \bar u(l',s_f)
\gamma^{\mu} u(l,s_l) \bar u(l,s_l) \gamma^{\nu}
u(l',s_f)
\nonumber \\
&=& {\rm Tr} \left [  (/\hspace{-1.7mm} l+\mu) \frac{1+
\gamma_5 /\hspace{-2.1mm} s_l}{2}
\gamma^{\nu} ( /\hspace{-1.7mm} l'+\mu) \gamma^{\mu}
 \right] .
\end{eqnarray}
Here $l$ and $l'$ are the initial and final lepton momenta, and $s_l$
is a spin vector of the initial lepton determined in (\ref{sp}).

The sum and difference of the cross sections with parallel and
antiparallel longitudinal polarization of a proton and a lepton
are expressed in terms of the spin-average and spin--dependent
hadron and  lepton tensors
\begin{equation}
\label{l+-}
{\cal L}^{\mu \nu}(\pm)=\frac{1}{2}({\cal L}^{\mu \nu}(+\frac{1}{2})
\pm {\cal L}^{\mu \nu}(-\frac{1}{2})),
\end{equation}
where ${\cal L}^{\mu;\nu}(\pm\frac{1}{2})$ are the tensors with the
helicity of the initial  lepton equal to $\pm 1/2$. The
tensors (\ref{l+-}) look like
\begin{eqnarray}
\label{lpm}
{\cal L}^{\mu \nu}(+)&=& 2 (g^{\mu \nu} l \cdot q + 2 l^\mu l^\nu -
l^\mu q^\nu - l^\nu q^\mu),\nonumber \\
{\cal L}^{\mu \nu}(-)&=& 2 i \mu \epsilon^{\mu\nu\delta\rho} q_{\delta}
(s_{l})_{\rho}.
\end{eqnarray}

\subsection{Pomeron--Proton Coupling}
The pomeron  is a vacuum $t$-channel exchange that describes
diffractive processes at high energies. The hadron-hadron
scattering amplitude, which is determined by the pomeron,
can be written in the form
\begin{equation}
\label{tpom}
T(s,t)={\rm i} I\hspace{-1.6mm}P(s,t)
V_{AI\hspace{-1.1mm}P} \otimes V^{BI\hspace{-1.1mm}P},
\end{equation}
where $I\hspace{-1.6mm}P$ is a function determined by the pomeron and
$V_{AI\hspace{-1.1mm}P}$ and $V^{BI\hspace{-1.1mm}P}$ are the
pomeron couplings with particles $A$ and $B$, respectively.

The spin structure of the pomeron coupling is an open problem
now. When the gluons from the pomeron couple to a single quark
in the hadron, a simple matrix structure of the pomeron vertex
\begin{equation}
\label{v_alf}
V_{h
I\hspace{-1.1mm}P}^{\alpha} =B_{h I\hspace{-1.1mm}P}
\gamma^{\alpha}
\end{equation}
appears. This standard coupling leads to spin-flip effects
decreasing with increasing energy like $1/s$. The effective
pomeron coupling with the hadron (\ref{v_alf}) is like a $C= +1$
isoscalar photon vertex \cite{lansh-m}. Then, the
pomeron--proton coupling should be equivalent to the isoscalar
electromagnetic nucleon current. The spin-dependent pomeron
coupling can be obtained if one  considers together with the
Dirac form factor (\ref{v_alf}) the Pauli form factor
\cite{nach} in the electromagnetic nucleon current. If the
gluons from the pomeron carries some fraction $x_P$ of the
initial proton momenta, this coupling can be written in the form
\begin{equation}
\label{ver}
V_{pgg}^{\alpha}(p,t,x_P)= 2 p^{\alpha}
A(t,x_P)+\gamma^{\alpha}  B(t,x_P).
\end{equation}

Let us study the meson-nucleon scattering which is described at
high--energies and fixed momentum transfer by the pomeron
contribution. The coupling of the pomeron with the meson for
small $|t|$ can be written in the form $q^\mu \phi(t)$ where $q$
is the initial meson 4-momentum and  $\phi(t)$ is a
meson-pomeron form factor. This form is similar to the
photon--meson vertex. For the pomeron--proton coupling
(\ref{ver}) we find the following meson-nucleon scattering
amplitude
\begin{equation}
\label{fmn}
M(\tilde{s},t)= i \left[\tilde{s} A(t,x_P)+/\hspace{-2.1mm} q B(t,x_P)
  \right]\phi(t).
\end{equation}
Here $\tilde{s}=(p+q)^2$ and $x_P \propto 1/\tilde{s}$. Thus,
the pomeron coupling (\ref{ver}) provides the standard form of
the scattering amplitude. The meson--proton helicity-non-flip
and helicity-flip amplitudes are expressed in terms of the
invariant functions $A$ and $B$
\begin{eqnarray}
\label{fnf}
F_{++}(s,t)&=& i \tilde{s} [B(t,x_P)+ 2 m A(t,x_P)] \phi(t);\nonumber\\
F_{+-}(s,t)&=& i \tilde{s} \sqrt{|t|} A(t,x_P) \phi(t),
\end{eqnarray}
and the spin-average cross-section is written in the form
\begin{equation}
\label{mp_sd}
\frac{d\sigma}{dt} \sim  [|B+2 m A|^2+ |t| |A|^2] \phi(t)^2.
\end{equation}
Here $m$ is a proton mass. We see that the term proportional to
$B$ represents the standard pomeron coupling that leads to the
non-flip amplitude. The $A$ function  is the spin--dependent
part of the pomeron coupling which produces  spin--flip effects
non vanishing at high-energies.  The absolute value of the
ratio of $A$ to $B$ is proportional to the ratio of
helicity-flip and non-flip amplitudes. It has been found in
\cite{gol_mod,gol_kr} that the ratio $|A|/|B| \sim 0.1 -0.2
\,\mbox{GeV}^{-1}$ and has a weak energy dependence. This value
of the $|A|/|B|$ ratio and weak energy dependence of spin
asymmetries in exclusive reactions is not in contradiction with
the experiment \cite{akch} and is confirmed by the model results
(see \cite{gol_mod,models} e.g.).

The proton-pomeron coupling similar to (\ref{ver}) has been used
in \cite{klen} to analyze the spin effects in diffractive
vector-meson production. In the model \cite{gol_mod}, the form
(\ref{ver}) was found to be valid for small momentum transfer
$|t| <  \mbox{few GeV}^2$ and the $A(t)$ function to be caused
by the meson-cloud effects in the nucleon. This model gives a
quantitative description of meson-nucleon and nucleon--nucleon
polarized scattering at high energies. The complicated spin
structure of the pomeron coupling can be due to the
nonperturbative structure of the proton. Really, in a QCD--based
model, in which the proton is viewed as being composed of a quark
and a diquark \cite{kroll}, the structure (\ref{ver}) for the
proton coupling with a two-gluon system  has been found for
moderate momentum transfer \cite{gol_kr}. The $A(t)$ contribution
is determined there by the effects of vector diquarks inside the
proton, which are of an order of $\alpha_s$. In all the cases
the spin-flip $A(t)$ contribution is determined by the
nonperturbative effects in the proton.

Since the pomeron consists of a two-gluon \cite{low}, the
pomeron coupling should have two gluon indices. We use in
calculations the following generalization of (\ref{ver}) as an
ansatz:
\begin{eqnarray}
\label{ver2}
V_{pgg}^{\alpha\alpha'}(p,t,x_P,l)&=& 4 p^{\alpha} p^{\alpha'}
A(t,x_P,l)\nonumber\\
&+&
(\gamma^{\alpha} p^{\alpha'}+\gamma^{\alpha'} p^{\alpha}) B(t,x_P,l).
\end{eqnarray}
This vertex is shown in Fig.1\ a,\ b by the blob in the proton
line. The properties of the structure (\ref{ver2}) are
completely equivalent to the coupling (\ref{ver}) (see Sect. 3.3
and 3.4). It has been mentioned above that this vertex contains
the gluon ladder, except the two gluons which provide the $l$
dependencies in (\ref{ver2}). In what follows we shall calculate
the imaginary part of the pomeron contribution to the scattering
amplitude which dominates in the high-energy region. This
contribution is equivalent to the $t$--channel cut in the
gluon--loop graph.

\subsection{Simple Form of Hadronic Tensor}
The hadronic tensor for the vertex (\ref{ver}) can be written
in the form
\begin{eqnarray}
\label{wstenz}
W^{\alpha;\beta}(s_p)&=& \sum_{spin \; s_f} \bar u(p',s_f)
 V_{pgg}^{\alpha}(p,t,x_P) u(p,s_p) \cdot \nonumber\\
&&\bar u(p,s_p)V_{pgg}^{\beta\,+}(p,t,x_P) u(p',s_f).
\end
{eqnarray}
Here $p$ and $p'$ are the initial and final proton momenta and
$s_p$ is a spin vector of the initial proton. The spin--average
and spin--dependent hadron tensors are determined similarly to
(\ref{l+-}). The leading term of the spin average hadron tensor
must have a maximum number of the large proton momenta
$p^{\alpha}$. It look like
\begin{equation}
\label{ws+f}
W^{\alpha;\beta}(+) = 4 p^{\alpha} p^{\beta}
( |B+2 m A|^2 + |t| |A|^2)
\end{equation}
and is proportional to the meson-proton cross section
(\ref{mp_sd}) up to some function of $t$.

The  spin-dependent part of the hadron tensor can be represented
as a sum of structures which have a different nature
\begin{equation}
\label{ws-f1}
W^{\alpha;\beta}(-) = \Delta
A_{full}^{\alpha;\beta} +
\Delta A_{1}^{\alpha;\beta}.
\end{equation}
 The $\Delta A_{full}$ term has  indices of
different pomeron couplings in the $\epsilon$ function.
\begin{equation}
\label{asf}
\Delta A_{full}^{\alpha;\beta}=2 i m |B|^2
 \epsilon^{\alpha\beta\gamma\delta}
(r_P)_\gamma (s_p)_\delta
\end{equation}
This contribution is proportional to $|B|^2$ and equivalent in
form to the spin-dependent part of the leptonic tensor (see
(\ref{lpm})) and called by us a full block asymmetry. The
$\Delta A_{1}$ term contains both $A$ and $B$ amplitudes from
(\ref{ver}):
\begin{eqnarray}
\label{das1}
\Delta A_{1}^{\alpha;\beta}&=&\left[4  p^{\beta}
B \right] \left[2 i A^{\star}   \epsilon^{\alpha\gamma\delta\rho}
p_{\gamma} (r_P)_{\delta} (s_p)_{\rho}
\right]\nonumber \\
&-&\left[4  p^{\alpha}
B^{\star} \right] \left[2 i A  \epsilon^{\beta\gamma\delta\rho}
p_{\gamma} (r_P)_{\delta} (s_p)_{\rho}
\right].
\end{eqnarray}

\subsection{Generalized Hadronic Tensor}
The hadronic tensor for the ansatz (\ref{ver2}) is given by
\begin{eqnarray}
\label{wtenz}
W^{\alpha\alpha';\beta\beta'}(s_p)&=& \sum_{spin \; s_f} \bar u(p',s_f)
 V_{pgg}^{\alpha\alpha'}(p,t,x_P,l) u(p,s_p)\cdot \nonumber\\
&&\bar u(p,s_p) V_{pgg}^{\beta\beta'\,+}(p,t,x_P,l') u(p',s_f).
\end{eqnarray}
The spin--average and spin--dependent hadron tensors are
written as
\begin{equation}
\label{wpm}
W^{\alpha\alpha';\beta\beta'}(\pm)=\frac{1}{2}
( W^{\alpha\alpha';\beta\beta'}(+\frac{1}{2})
 \pm W^{\alpha\alpha';\beta\beta'}(-\frac{1}{2})).
\end{equation}
For the leading term of $W(+)$ we find
\begin{equation}
\label{w+f}
W^{\alpha\alpha';\beta\beta'}(+) = 16 p^{\alpha} p^{\alpha'}
p^{\beta}  p^{\beta'} ( |B+2 m A|^2 + |t| |A|^2).
\end{equation}
Note that we omit for simplicity here and in what follows the
arguments of the $A$ and $B$ functions unless it is necessary.
However, we shall remember that the amplitudes $A$ and $B$
depend on $l$, otherwise the complex conjugative values
$A^\star$ and $B^\star$ are the functions of $l'$.

The  spin-dependent part of the hadron tensor can be written as
\begin{equation}
\label{w-f1}
W^{\alpha\alpha';\beta\beta'}(-) = \Delta
A_{full}^{\alpha\alpha';\beta\beta'} +
\Delta A_{1}^{\alpha\alpha';\beta\beta'}.
\end{equation}
Here
\begin{eqnarray}
\label{af}
&&\Delta A_{full}^{\alpha\alpha';\beta\beta'}=2 i m |B|^2\cdot \nonumber\\
&&\left[
p^{\alpha'} p^{\beta'} \epsilon^{\alpha\beta\gamma\delta}
(r_P)_\gamma (s_p)_\delta
 +\left(^{\mbox{All}\ \
\mbox{Per-}} _{\mbox{mutations}}\right)
\left(^{\{\alpha \to \alpha' \}} _{\{\beta \to \beta' \}}\right)
 \right]
\end{eqnarray}
and
\begin{eqnarray}
\label{da1}
&&\Delta A_{1}^{\alpha\alpha';\beta\beta'}=\nonumber\\
&&\left[4  p^{\beta}  p^{\beta'}
B \right] \left[2 i A^{\star}  p^{\alpha'} \epsilon^{\alpha\gamma\delta\rho}
p_{\gamma} (r_P)_{\delta} (s_p)_{\rho} +\{ \alpha \to \alpha'\}
\right]- \nonumber \\
&&\left[4  p^{\alpha}  p^{\alpha'}
B^{\star} \right] \left[2 i A  p^{\beta'} \epsilon^{\beta\gamma\delta\rho}
p_{\gamma} (r_P)_{\delta} (s_p)_{\rho} +\{ \beta \to \beta'\}
\right].
\end{eqnarray}

 The $\Delta A_{1}$ term has a form of a product of
the spin-dependent part $\Delta A$ of one proton vertex to the
symmetric part $S$ of the other. Really, it can be written as
\begin{equation}
\label{a1}
\Delta A_{1}= -(\Delta A^{\alpha\alpha'})^{\star}
S^{\beta\beta'}- \Delta A^{\beta\beta'}
(S^{\alpha\alpha'})^{\star},
\end{equation}
where
\begin{eqnarray}
\label{sa}
S^{\beta\beta'} &=&\left  [4 B p^{\beta} p^{\beta'}\right]; \nonumber \\
 \Delta A^{\alpha\alpha'} &=& \left [2 i
A p^{\alpha'} \epsilon^{\alpha\gamma\delta\rho}
p_{\gamma} (r_P)_{\delta} (s_p)_{\rho} +\{ \alpha \to \alpha'\}\right
].
\end{eqnarray}

We see that $\Delta A_{full}$ and $\Delta A_{1}$ from (\ref{af})
in contrast with the relevant terms in (\ref{asf}) have the
additional $p^{\alpha'}  p^{\beta'}$ momenta and symmetrization
over  $ \alpha \to \alpha'$, $\beta \to \beta'$  indices. The
powers of large scalar production  $p \cdot q$ which appear
in this case will be compensated after the loop integration over
$l$ and $l'$. As a result, (\ref{ws-f1}) and (\ref{w-f1}) will
produce the same spin-dependent amplitude. Using the mentioned
argument we find that the forms (\ref{ws+f}) and (\ref{w+f})
lead to the same spin--average amplitude too. Hence, the pomeron
couplings (\ref{ver}) and (\ref{ver2}) are equivalent. The very
important property of (\ref{ws-f1},\ref{w-f1}) is that both the
$B^2$ and $A \cdot B$ terms contribute to the $W(-)$ tensor
which is responsible for the asymmetry.

\section{Diffractive $J/\Psi$ Leptoproduction}\label{sect4}
\subsection{Amplitude of the $\gamma  I\hspace{-1.8mm}P \to J/\Psi$
Transition}
 Now we are passing to the structure of the amplitude
of the $\gamma I\hspace{-1.6mm}P \to J/\Psi$ production. In what
follows we have regarded the $J/\Psi$ meson as an $S$-wave system
of $c \bar c$ quarks \cite{berger}. The $J/\Psi$-wave function
in this case has a form $g (/\hspace*{-0.20cm} k+m_c)
\gamma_\mu$ where $k$ is the momentum of quark and $m_c$ is its
mass. In the nonrelativistic approximation both the quarks have
the same momenta $k$ equal to half of the vector meson momentum
$K_J$ and the mass of $c$ quark is equal to $m_J/2$. The
coupling constant $g$ can be expressed through the $e^+ e^-$
decay width of the  $J/\Psi$ meson
\begin{equation}
g^2=\frac{3 \Gamma^J_{e^+ e^-} m_J}{64 \pi \alpha^2}.
\end{equation}

 The gluons from the pomeron are coupled with the single and
different quarks in the $c \bar c$ loop (see Fig. 1 a, b). The
$\gamma  I\hspace{-1.6mm}P \to J/\Psi$ transition amplitude for
these graphs look like
\begin{eqnarray}
\label{tr_j}
T_a&=& g \mbox{Tr}[/\hspace*{-0.20cm} e_J (/\hspace*{-0.20cm} k+m_c)
\gamma_{\alpha} (/\hspace*{-0.20cm} k+/\hspace*{-0.20cm} l+m_c)
 \gamma_{\alpha'} (/\hspace*{-0.20cm} r+m_c) \gamma_{\nu}]
\frac{1}{r^2-m_c^2};\nonumber\\
T_b&=&  g \mbox{Tr}[/\hspace*{-0.20cm} e_J (/\hspace*{-0.20cm} k+m_c )
\gamma_{\alpha'}  (/\hspace*{-0.20cm} w + m_c)
 \gamma_{\nu} (-/\hspace*{-0.20cm} k - /\hspace*{-0.20cm} l+m_c)
\gamma_{\alpha}]\frac{1}{w^2-m_c^2}.
\end{eqnarray}
Here $r=k-r_P$ and $w=k-r_P-l$ are the momenta of the
off-mass-shell quark in the loop for the diagram, Fig 1.\ a, b,
respectively and $e_J$ is  polarization of the $J/\Psi$ meson
which obeys the relation
\begin{equation}
\label{e_j}
\sum_{Spin_J} e_J^{\rho} (e_J^{\sigma})^+=-g^{\rho \sigma}+
\frac{K_J^{\rho} K_J^{\sigma}}{m_J^2}.
\end{equation}

 It is known (see \cite{rys,j-psi} e.g) that the leading terms
of the amplitude of the diffractive vector meson production is
mainly imaginary. To calculate it we must consider the
$\delta$-function contribution in the $s$-channel propagators
($k+l$ and $p'-l$ lines for  Fig 1.\ a). With the help of these
$\delta$ functions  the integration over $l$
\begin{equation}
\int d^4 l= \frac{1}{2} \int d l_{+}  d l_{-}  d l_{\perp}
\end{equation}
can be carried out over $l_{+}$ and $l_{-}$ variables. One can
find that both the $l_{\pm}$ components of the vector $l$ are
small: $l_{+} \sim l_{-} \propto 1/p_{+}.$ This results in the
transversity of the gluon momentum $l^2 \simeq -l_{\bot}^2$. The
same is true for integration over $l$ in the nonplanar graph of
Fig 1.\ b. For the arguments in the propagator of  graphs, Fig
1.\ a, b, we find
\begin{eqnarray}
\label{w}
r^2-m_c^2 &=& -\frac{M_J^2+Q^2+|t|}{2}, \nonumber\\
w^2-m_c^2 &=& -2 \left(l_{\bot}^2 + \vec l_{\bot} \vec r_{\bot}
+\frac{M_J^2+Q^2+|t|}{4} \right).
\end{eqnarray}
Thus these quark lines are far from the mass shell for heavy
vector meson production even for small $Q^2$ \cite{rys}.

\subsection{Cross Section of Vector Meson Production}
The spin-average and spin dependent cross
sections with parallel and antiparallel longitudinal
polarization of a lepton and a proton are determined by the
relation
\begin{equation}
\label{spm}
\sigma(\pm) =\frac{1}{2} \left( \sigma(^{\rightarrow}
_{\Leftarrow}) \pm \sigma(^{\rightarrow} _{\Rightarrow})\right).
\end{equation}
These cross sections are expressed through the squared amplitude
of the $\gamma  I\hspace{-1.6mm}P \to J/\Psi$ transition
convoluted with the spin-average and spin dependent
lepton and hadron tensors (\ref{lpm}),  (\ref{wpm}-\ref{w-f1}).
The analyses of the leading over $s$ contribution to the cross
sections have been carried out with the help of the REDUCE and MAPLE
programs. We summarize over the spin of the $J/\Psi$ meson and use
(\ref{e_j}) in calculation. In both the cases the squared
amplitude of the $J/\Psi$ electroproduction is expressed through the
integral over $l, l'$
\begin{eqnarray}
\label{tpm}
|T^{\pm}|^2 &=& \int d^2 l_{\bot} d^2 l_{\bot}' D(t,Q^2,l_{\bot})
D(t,Q^2,l'_{\bot})\cdot \nonumber\\
&& F^{\pm}[A(l_{\bot}),B(l_{\bot});
A^{\star}(l'_{\bot}),B^{\star}(l'_{\bot})],
\end{eqnarray}
where the functions $F^{\pm}$ include the $A$ and $B$ amplitudes
from (\ref{ver2}). The $l$ dependence of these functions for
small $l$ has been discussed, e.g., in the second Ref. of
\cite{cudell}. The function $D$ is determined by the contribution
of the $t$-channel gluon propagators and the sum of the  $\gamma
I\hspace{-1.6mm}P \to J/\Psi$ transition amplitude (\ref{tr_j})
for the graphs of Fig.\ 1a, b
\begin{eqnarray}
\label{d}
D(t,Q^2,l_{\bot}) &=&\frac{1}{(l_\perp^2+\lambda^2)
((\vec l_\perp+\vec r_{\perp})^2+\lambda^2)}\cdot\nonumber\\
&&
\left(\frac{n_a}{r^2-m_c^2} + \frac{n_b}{w^2-m_c^2}  \right).
\end{eqnarray}
The leading over $s$ terms in the numerators $n_{a(b)}$ for the
graphs, Fig.\ 1a, b,  have a similar form but they  are different
in sign: $n_b \sim -n_a =n$. In the sum of diagrams, Fig.\ 1a, b,
their contributions mainly compensate each other:
\begin{eqnarray}
\label{prop}
&&\frac{n}{w^2-m_c^2}-\frac{n}{r^2-m_c^2}=\nonumber\\
&&\frac{2 n (l_{\bot}^2 + \vec l_{\bot} \vec r_{\bot})}
{\left( M_J^2+Q^2+|t| \right)  \left[l_{\bot}^2 +
\vec l_{\bot} \vec r_{\bot}
+(M_J^2+Q^2+|t|)/4 \right]}.
\end{eqnarray}
This function determines the $Q^2$-dependence of $D$.  It can be
seen that the typical scale in the integral (\ref{tpm}) is defined
by (\ref{prop}) and of the order $\bar l_{\bot}^2 \sim
(M_J^2+Q^2+|t|)/4$ \cite{rys,j-psi}. As a result, (\ref{tpm})
can be estimated in the form
\begin{eqnarray}
\label{test}
|T^{\pm}|^2 &=& F^{\pm}(A,B;A^{\star},B^{\star})
I^2;\nonumber\\
I&=& \int d^2 l_{\bot} D(t,Q^2,l_{\bot}).
\end{eqnarray}
The functions $A$ and $ B$ in $ F^{\pm}$ are dependent on the
scale
\begin{equation}
\label{qscale}
\bar l_{\bot}^2 \sim \bar Q^2=(M_J^2+Q^2+|t|)/4.
\end{equation}

The cross section of the $J/\Psi$ leptoproduction can be written in
the form
\begin{equation}
\label{ds_tqy}
\frac{d\sigma^{\pm}}{dQ^2 dy dt}=\frac{|T^{\pm}|^2}{32 (2\pi)^3
 Q^2 s^2 y}.
\end{equation}
For the spin-average squared amplitude we find
\begin{eqnarray}
\label{t+}
|T^{+}|^2 &=&  N ((2-2 y+y^2) m_J^2 + 2(1 -y) Q^2) s^2 \cdot \nonumber\\
&&
[|B+2 m A|^2+|A|^2 |t|] I^2.
\end{eqnarray}
Here $N$ is a normalization factor
\begin{equation}
\label{n}
N=\frac{\Gamma^J_{e^+ e^-} M_J \alpha_s^4 }{27 \pi^2}.
\end{equation}
In (\ref{t+}) the term proportional to $(2-2 y+y^2) m_J^2$
represents the contribution of a virtual photon  with transverse
polarization. The $2(1 -y) Q^2$ term describes the effect of
longitudinal photons. Thus, we see that the ratio of
$\sigma_T/\sigma_L \propto m_J^2/Q^2$. Such a behaviour is typical
of a simple form of the vector meson wave function used here
(see e.g. \cite{cudell_t})

The spin-dependent squared amplitude look like
\begin{equation}
\label{t-}
 |T^{-}|^2= N (2- y)  s |t|
[|B|^2+ m (A^\star B +A B^\star)] m_J^2 I^2.
\end{equation}

We see that $|T^{-}|^2$ vanishes in the forward direction
($t=0$) and is suppressed as a power of $s$ with respect to
(\ref{t+}). The reason for this suppression is quite simple. The
leading contribution to $\sigma(-)$ is going from the term
$\epsilon^{\alpha\beta\gamma\rho}(r_P)_{\gamma}..$ which is
proportional  to $x_{P} p$. As a result, an additional $x_p$
appears in $\sigma(-)$. It has been confirmed by the
calculation of the $A_{ll}$ asymmetry in different diffractive
reactions \cite{gol_jpsi,gol_all}. In the case of vector meson
production, $x_P$ is small (\ref{x_P}) and behaves like $1/s$.
Hence, longitudinal double--spin asymmetry in this diffractive
process will be small at high energies which is confirmed by our
calculation for (\ref{t-}) and (\ref{t+}).

\section{Numerical Results for Spin Dependent Cross Section}\label{sect5}
We shall calculate the polarized cross section of the diffractive
$J/\Psi$ production (\ref{ds_tqy}) for the amplitudes (\ref{t+},
\ref{t-}).
The connection of the spin-average cross section of the $J/\Psi$
production with the gluon distribution function is  known
\cite{rys,j-psi}
\begin{equation}
\label{dsg}
\left. \frac{d \sigma}{dt}\right|_{t \sim 0} \propto
F_B^2(t) \left(x_P G(x_P,\bar Q^2)  \right)^2.
\end{equation}
Here $F_B(t)$ is a new form factor which describes
$t$--dependence of the two-gluon coupling with the proton. The
expression of this cross section through the pomeron-proton
structure has been found in (\ref{t+}). It can be seen that the
$B$ function in (\ref{ver2}) can be written as a product of the
form factor and the gluon distribution
\begin{equation}
\label{b_g}
B(t,x_P, \bar Q^2) =F_B(t) \left( x_P G(x_P,\bar Q^2)  \right).
\end{equation}
 As the pomeron--proton vertex might be similar to the photon-proton
coupling \cite{mod}, we shall use a simple approximation
\begin{equation}
\label{fb}
F_B (t) \sim F^{em}_p(t),
\end{equation}
where $F^{em}_p(t)$ is the standard form for the electromagnetic
form factor of the proton
\begin{equation}
\label{fp}
F^{em}_p(t)=\frac{(4 m_p^2+2.8 |t|)}{(4 m_p^2+|t|)(
1+|t|/0.7GeV^2)^2}.
\end{equation}

It has been shown in (\ref{t+}, \ref{t-}) that the leading
contribution to the $T^+$ and $T^-$ amplitudes is determined by
the same loop integral $I$. For simplicity we shall suppose that
the $A$ amplitude can be parameterized in the form similar to
(\ref{b_g})
\begin{equation}
\label{a_g}
A(t,x_P, \bar Q^2) = \alpha F_A(t) \left(x_P G(x_P,\bar Q^2)
\right).
\end{equation}
As previously, the new form factor $F_A(t)$ describes the
$t$--dependence of the two-gluon coupling with the proton for
the $A$ function. We shall use for simplicity $ F_A=F^{em}_p$.
For the approximation (\ref{a_g}), the ratio of the $A$ and $B$
amplitudes is independent of $x$. The $\alpha=A/B$ ratio
determines through the $x$ dependences of the functions ($x \sim
1/s$ in this case)  the energy behaviour of the spin asymmetries
in exclusive reactions at high energies and fixed momentum
transfer. Thus, (\ref{b_g}) and (\ref{a_g}) result in energy
independence of spin  asymmetries which is in agreement with
their weak energy dependence obtained in the models
\cite{gol_mod,gol_kr,models}. To study the $\alpha$ sensitivity
of the cross section we shall use in our estimations the  value
of $|\alpha|\leq 0.1\mbox{GeV}^{-1}$. It has been mentioned above
that this magnitude is consistent with the model estimations
\cite{gol_mod,gol_kr}.

The energy dependence of the cross sections is determined by the
pomeron contribution to the gluon distribution function at small
$x$
\begin{equation}
\label{g_x}
\left(x_P G(x_P,\bar Q^2) \right) \sim
\left( \frac{s y}{m_J^2+Q^2+|t|} \right) ^{(\alpha_p(t)-1)}.
\end{equation}
Here $\alpha_p(t)$ is a pomeron trajectory. The linear
approximation of the pomeron trajectory (\ref{pomer}) is used.
The parameters of this trajectory was determined from the fit of
the diffractive $J/\Psi$ production by ZEUS \cite{zeus_p}
\begin{equation}
\label{z_p}
\alpha_p(t)=1+(0.175 \pm 0.026)+ (0.015 \pm 0.065) t.
\end{equation}

In our model estimations of the polarized cross section of the
diffractive $J/\Psi$ production we shall use the values
$\epsilon=0.15$ and $ \alpha'=0$ which are not far from
(\ref{z_p}). The typical scale of the reaction is determined by
(\ref{qscale}). For not large $Q^2$ and $|t|$  the value of
$\bar Q^2$ is about 2.5-3.0 $\mbox{GeV}^2$. In this region we
can work with fixed  $\alpha_s \sim 0.3$. An effective gluon
mass in (\ref{d}) is chosen to be equal to 0.3 $\mbox{Gev}^2$. The
cross section depends on this parameter weakly. The value of
$\Gamma^J_{e^+ e^-}=5.26 \mbox{keV}$ is used.

We shall integrate the cross sections (\ref{ds_tqy}) over $Q^2$ and
$y$ to get the differential cross sections of the $J/\Psi$
production
\begin{equation}
\label{ds_qyint}
\frac{d\sigma^{\pm}}{dt}=\int_{y_{min}}^{y_{max}}dy
\int_{Q^2_{min}}^{Q^2_{max}} dQ^2
\frac{d\sigma^{\pm}}{dQ^2 dy dt},
\end{equation}
where $$ Q^2_{min}=m_e^2\frac{y^2}{1-y}\;\; \mbox{and}\;\;
Q^2_{max}=4\mbox{Gev}^2. $$

\begin{figure}
\epsfxsize=8cm
\centerline{\epsfbox{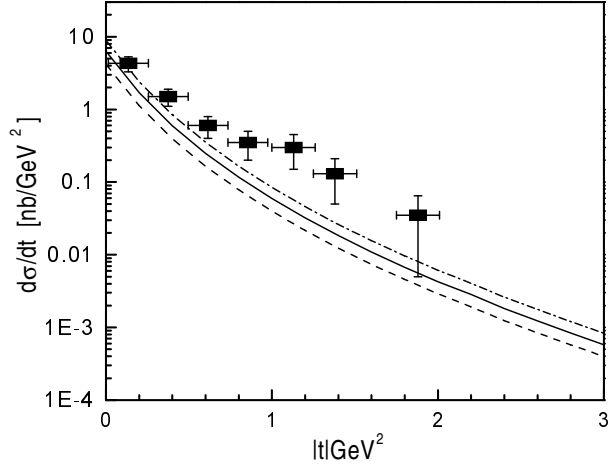}}
\caption{The differential cross section of the $J/\Psi$ production
at HERA energy: solid line -for $\alpha=0$; dot-dashed line -for
$\alpha=0.1\mbox{GeV}^{-1}$;
dashed line -for $\alpha=-0.1\mbox{GeV}^{-1}$. Data are from
\cite{jpsi1}.}
\label{fig:2}       
\end{figure}
For the HERA energy $\sqrt{s}=300\mbox{GeV}$ we  integrate over
the energy in the photon--proton system 30 GeV $ < W_{\gamma p}
<$ 150 GeV ($W_{\gamma p} \sim \sqrt{y s}$) which is equivalent
to the range 0.01 $<y<$ 0.25. Our results,  shown in Fig.\ 2,
describe  experimental data \cite{jpsi1} at small $|t|$ and lie
below them a little for the momentum transfer larger than 1
GeV$^2$. This may indicate that the simple approximation of the
form factor (\ref{fb}) used here is not good for $|t| > 1
\mbox{GeV}^2$ . Our estimation for the HERMES energy
$s=50\mbox{GeV}^2$ is performed for integration over $0.3 < y <
0.7$. The predicted cross sections are shown in Fig.\ 3. It is
seen from  Figs.\ 2 and 3 that the spin-average cross sections
are sensitive to $\alpha$ but the shape of all curves are the
same. Thus, it is difficult to extract information about the
spin--dependent part of the pomeron coupling from the
spin--average cross section of the diffractive vector meson
production.
\begin{figure}
\epsfxsize=8cm
\centerline{\epsfbox{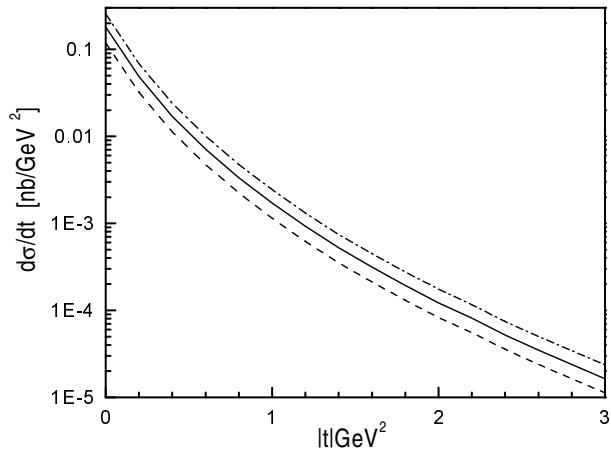}}
\caption{ The differential cross section of the $J/\Psi$ production
at HERMES energy: solid line -for $\alpha=0$; dot-dashed line -for
$\alpha=0.1\mbox{GeV}^{-1}$;
dashed line -for $\alpha=-0.1\mbox{GeV}^{-1}$.}
\label{fig:3}       
\end{figure}

Using the same formulae we  calculate the cross section
$\sigma(-)$. This gives us a possibility to estimate the
longitudinal double--spin $A_{ll}$ asymmetry of the $J/\Psi$
production at high energies. As has been found, the asymmetry
vanishes as $1/s$ and for the HERA energy range the expected
value of $A_{ll}$ will be negligible. The predicted asymmetry
for  HERMES as a function $\alpha$ is shown in Fig.\ 4. The
important property of $A_{ll}$  is that the asymmetry of the
vector meson production is equal to zero in the forward
direction. The $A_{ll}$ asymmetry might be connected with the
spin-dependent gluon distribution $\Delta G$ only for $|t|=0$.
Thus, $\Delta G$ cannot be extracted from $A_{ll}$ in agreement
with the results of \cite{mank}.
\begin{figure}
\epsfxsize=8cm
\centerline{\epsfbox{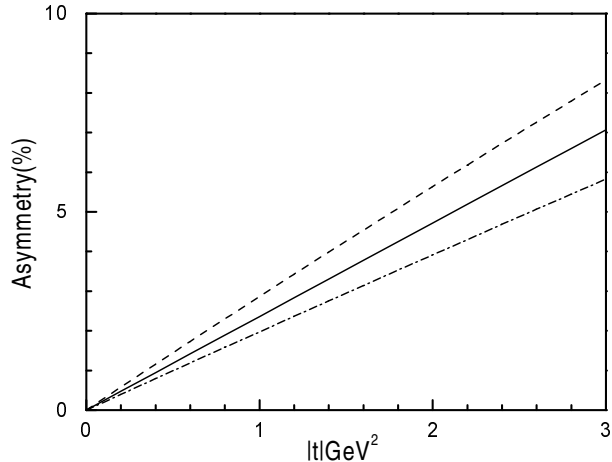}}
\caption{The predicted $A_{ll}$ asymmetry of the $J/\Psi$ production
at HERMES: solid line -for $\alpha=0$; dot-dashed line -for
$\alpha=0.1\mbox{GeV}^{-1}$;
dashed line -for $\alpha=-0.1\mbox{GeV}^{-1}$.}
\label{fig:4}       
\end{figure}

To understand the $\alpha$ dependence of the asymmetry we
shall use the approximated expression of the integral
(\ref{ds_qyint}). The functions $d\sigma^{\pm}/(dQ^2 dy dt)$
decrease very rapidly with growing $Q^2$. Calculating only the
leading Log terms of the integral over $Q^2$ in (\ref{ds_qyint})
we can write the double spin asymmetry $A_{ll}$ in a simple form
\begin{equation}
\label{all_a}
A_{ll} \sim \frac{|t|}{s}\frac{(2-\bar y)(1+2 m \alpha)}
{(2-2\bar y+\bar y^2)\left[ (1+2 m \alpha)^2+\alpha^2|t| \right]
},
\end{equation}
where $\bar y$ is some average value in the integration region.
We find, that the asymmetry is non zero for $\alpha=0$
\begin{equation}
\label{all0}
A_{ll}^0=A_{ll}(\alpha=0)
\simeq \frac{|t|}{s}\frac{(2-\bar y)}
{(2-2\bar y+\bar y^2)}.
\end{equation}
 This term in asymmetry is determined by the $\Delta
A_{full}$ contribution to $W(-)$ (see (\ref{af})). Both
spin-average and spin-dependent cross sections (\ref{t+},
\ref{t-}) are proportional to $|B|^2 \sim \left(x_P G(x_P,\bar
Q^2) \right)$ for $\alpha=0$, which results in the independence
of $A_{ll}^0$ on the gluon distribution. We see from Fig.\ 4
that the  $A_{ll}^0$ part of the asymmetry dominates. The value
of the asymmetry for $\alpha \neq 0$ is determined by the
spin--dependent part of the pomeron coupling. However, the
sensitivity of the asymmetry to $\alpha$ is not very strong.
Thus, it will not be so easy to study the spin structure of the
pomeron coupling with the proton from the $A_{ll}$ asymmetry of
the diffractive $J/\Psi$ production.

\section{Conclusion}\label{sect6}
In the present paper, the polarized cross section of the
diffractive $J/\Psi$ leptoproduction at high energies has been
studied. The relevant cross section can be determined in terms
of the leptonic and hadronic tensors; and the squared amplitude
of the vector meson production, through the photon-pomeron
fusion. The amplitude of the $\gamma  I\hspace{-1.6mm}P \to
J/\Psi$ transition is described by the simple non-relativistic
wave function. This approximation is efficient, at least for
heavy meson production. The introduced hadronic tensor is
expressed in terms of the pomeron-proton coupling structure
which has the helicity flip part. As a result, connection of the
spin--dependent cross section in the diffractive $J/\Psi$
production with the pomeron coupling has been found. We predict
the not small value of the $A_{ll}$ asymmetry of the diffractive
vector meson production at the HERMES energy. However, the
asymmetry decreases as $1/s$ with growing energy and at the HERA
energy it will be extremely small. It  has been found that the
$A_{ll}$ asymmetry  vanishes at $t=0$. Thus, it is impossible to
extract the polarized gluon distribution $\Delta G$ from the
asymmetry. The predicted asymmetry is independent of the mass of
a produced meson. We can expect a similar value of the asymmetry
in the polarized diffractive $\phi$ --meson leptoproduction.

The longitudinal double spin asymmetry of the vector meson
production for nonzero momentum transfer has been found to be
dependent on the $A$ term of the pomeron coupling which produces
 helicity--flip effects. Note that this spin-dependent part
of the coupling $A$ is parametrized here by the gluon structure
function of the proton $G$, for simplicity. Generally, the
function $A$ should be determined by the polarized gluon
distribution and the ratio $\alpha$ might depend on $x_P$ and
$t$.  However, this conformity is not known quite well now. To
find the explicit connection of $A$ with spin--dependent
gluon distribution in QCD, additional study is necessary. Our
results show the essential role of the " full block asymmetry"
in $A_{ll}$. This contribution does not depend on the gluon
distribution and has a kinematic character to all appearance.
The information about the spin-structure of the pomeron coupling
can generally be extracted from the $A_{ll}$ asymmetry of the
vector meson production for $|t| \neq 0$. Such investigations
can be carried out in future polarized experiments at CERN and
DESY. However, the sensitivity of  asymmetry to the
$\alpha$--ratio  is quite weak. Thus, the diffractive vector
meson production might not be a good tool to study the polarized
gluon distributions of the proton and spin structure of the
pomeron.

\section{Acknowledgments}
We would like to thank A. De Roeck, A. Efremov, P. Kroll, T. Morii,
O. Nachtmann, W.-D. Nowak and O. Teryaev for fruitful discussions.
This work was supported in part by the Heisenberg-Landau Grant.

\newpage

\end{document}